\documentclass{aa}
\usepackage{epsfig,lscape} 
\usepackage{rotating}

\def \rsun {\ifmmode$R$_{\odot}\else R$_{\odot}$\fi}

\def \hcm {\hbox {\ifmmode $ atoms cm$^{-2}\else atoms cm$^{-2}$\fi}}
\def\approxgt{\mathrel{\hbox{\rlap{\lower.55ex \hbox {$\sim$}}
        \kern-.3em \raise.4ex \hbox{$>$}}}}
\def\approxlt{\mathrel{\hbox{\rlap{\lower.55ex \hbox {$\sim$}}
\kern-.3em \raise.4ex \hbox{$<$}}}}
\include{epsf}

\begin{document}
 

\thesaurus{08.02.1,08.02.1,08.09.2,08.14.1,13.25.5}

\title{A BeppoSAX observation of the massive X-ray binary 4U\,1700-37}

\author{A.P. Reynolds\inst{1} \and A. Owens\inst{1}
\and L. Kaper\inst{2}
\and A.N. Parmar\inst{1}
\and A. Segreto\inst{3}
}

\institute{Astrophysics Division, Space Science Department of ESA,
ESTEC, P.O. Box 299, 2200 AG Noordwijk, The Netherlands
\and
Astronomical Institute ``Anton Pannekoek'', Kruislaan 403,
1098 SJ Amsterdam, The Netherlands
\and
Instituto Fisica Cosmica e Applicazioni all'Informatica, CNR, via La Malfa
153, 90146 Palermo, Italy
}

\offprints{A.P. Reynolds: areynold@astro.estec .esa.nl}
\date{Received ; accepted}

\maketitle

\markboth{A.P. Reynolds et al.}{4U\,1700-37}

\begin{abstract}
 
A 0.5--200~keV BeppoSAX spectrum of the non-pulsating high-mass X-ray binary 
4U\,1700-37 is presented.
The spectrum is well characterized by the standard
accreting pulsar model of an absorbed power-law with a photon
index of $1.07 \pm ^{0.02} _{0.03}$, sharply modified by an
exponential cutoff above $5.9 \pm 0.2$~keV. The e-folding energy
of the cutoff is $23.9 \pm 0.5$~keV. 
A soft bremsstrahlung component with a temperature of $0.2 \pm 0.1$~keV
is also required, together with a narrow iron line at 6.5~keV.
Both continuum components
are absorbed by a column of $(5.1 \pm 0.2)\times 10^{22}$~atom~cm$^{-2}$.
There is some evidence for the presence of a broad
cyclotron absorption feature
at $\sim$37~keV, although we cannot exclude the possibility that this
is due to an incorrect modeling of the continuum, or instrumental effects.
The hypothesis that the compact object is a
neutron star rather than a black hole seems most likely.

\end{abstract}

\section{Introduction}

4U\,1700-37 is a massive eclipsing X-ray binary 
consisting of a compact accreting 
object embedded in the wind of HD153919, a supergiant O star 
(Jones et al. 1973). Orbiting every 3.4 days, 
the compact object has an X-ray spectrum reminiscent of an accreting
highly magnetized neutron
star system (Haberl et al. 1989), although 
no X-ray pulsations have ever been confirmed. The X-ray flux is
significantly less than expected on the basis of standard
wind accretion theory.

The most recent estimates of the mass of HD~153919 and its
companion are from Heap \& Corcoran (1992), and Rubin et al. (1996).
Heap \& Corcoran (1992) propose a mass for HD~153919
of ${\rm 50 \pm 2 M_{\odot}}$, with a 
companion mass of ${\rm 1.8 \pm 0.4 M_{\odot}}$. 
Rubin et al. (1996)
reanalyzed the system parameters using Monte Carlo methods, finding
masses of 30$^{+11}_{-7}$ M$_{\odot}$ and 2.6 
$^{+2.3}_{-1.4}$ 
M$_{\odot}$ for HD~153919 and its companion, respectively. Both estimates
are, at their lower ranges, consistent with the compact object being 
a neutron star, but a black hole cannot be excluded.

Since pulsations have never been detected, further 
evidence for the object's
nature must be sought from its X-ray spectrum. 
In general, X-ray
pulsar spectra require an exponential cutoff to the continuum
between $\sim$10--20~keV, unlike black hole spectra, which
require either a single power-law or an ultra-soft component plus
hard power-law tail (e.g., Tanaka \& Lewin 1995).
Additionally, about half of the sample of 23 X-ray pulsars
observed by {\it Ginga} show cyclotron features in their spectra, usually
at energies between 10--60~keV (Makishima \& Mihara 1992; Mihara 1995),
indicating the presence of strong magnetic fields.

\section{Previous X-ray spectroscopy of 4U\,1700-37}

The X-ray spectrum of 4U\,1700-37 has been studied using many
different satellites, including HEAO-1 and {\it Einstein} (White et al. 1983),
and EXOSAT (Haberl et al. 1989). The spectral 
shape is highly reminiscent of other X-ray pulsars 
(White et al. 1983), being
well described by a power-law with photon
index, $\alpha$, of $\sim$0.15, 
modified by a high-energy cutoff above 6--7~keV.
Haberl \& Day (1992) confirmed the above results with {\it Ginga} observations,
also finding that $\alpha$ varies
between $-$0.5 and 1.0.  Haberl et al. (1994)
show that the 0.1--2.4~keV
ROSAT Position Sensitive Proportional Counter (PSPC)
spectrum is consistent with
a thermal bremsstrahlung model with a temperature, kT, fixed at 0.74~keV, as
originally found by {\it Ginga} (Haberl \& Day 1992), plus a hydrogen 
column density of $3.6 \times 10^{22}$~atom~cm$^{-2}$. 
Haberl \& Day (1992) also
found that the best-fit values of $\alpha$ 
varied depending on the photoelectric column
density. Since this is physically unrealistic, the authors show that
the spectral variations could also be modeled using a two-component
scattering model, consisting of two power-laws with the same slopes but
variable absorption, one of which represents a less absorbed scattered
component. 

The spectrum above 20 keV was examined using BATSE data 
by Rubin et al. (1996, and
references therein)
who confirmed earlier findings that it can be 
represented by a thermal 
bremsstrahlung model with kT $\sim$25~keV, 
out to 120~keV. There have been no reported
changes in the shape of the high energy spectrum 
due to orbital phase or source intensity.

While the above observations 
are individually useful, the launch of BeppoSAX, with its very wide 
spectral bandpass, offers the
first opportunity to study the entire X-ray spectrum of 4U\,1700-37 
simultaneously. BeppoSAX is proving to be a highly capable mission for
discovering and studying cyclotron lines (e.g., Dal Fiume et al. 1998).
For the first time, therefore, we present a single analysis of the entire
spectrum, obtained at one epoch, spanning 0.5--200~keV.

\section{Observations}

The X-ray astronomy satellite BeppoSAX (Boella et al. 1997a) contains 
four coaligned Narrow Field Instruments, or NFI. 
Results from the Low-Energy Concentrator Spectrometer (LECS;
0.1--10~keV; Parmar et al. 1997), Medium-Energy Concentrator
Spectrometer (MECS; 1.3--10~keV; Boella et al. 1997b),
High Pressure Gas Scintillation Proportional Counter
(HPGSPC; 5--120~keV; Manzo et al. 1997) and the Phoswich
Detection System (PDS; 15--300~keV; Frontera et al. 1997) 
are presented here. 
The MECS consisted of three identical grazing incidence
telescopes with imaging gas scintillation proportional counters in
their focal planes. The LECS uses an identical concentrator system as
the MECS, but utilizes an ultra-thin entrance window and
a driftless configuration to extend the low-energy response to
0.1~keV. The non-imaging HPGSPC consists of a single unit with a collimator
that is alternatively rocked on- and off-source. The non-imaging
PDS consists of four independent units arranged in pairs each having a
separate collimator. Each collimator can be alternatively
rocked on- and off-source.
 
4U\,1700-37 was observed by BeppoSAX between 1997 April 1 11:21 and
22:15~UTC. This interval corresponds to orbital phases 0.44--0.58, 
where mid-eclipse of the X-ray source occurs at phase 0.0, using
the ephemeris of Rubin et al. (1996). 
Good data were selected from intervals when the elevation angle
above the Earth's limb was $>$$4^{\circ}$ and when the instrument
configurations were nominal, using the SAXDAS 1.3.0 data analysis package.
The standard collimator dwell time of 96~s for each on- and
off-source position was used, together with rocking angles of 180\arcmin\ 
and 210\arcmin\ for the HPGSPC and PDS, respectively. 
The exposures 
in the LECS, MECS, HPGSPC, and PDS instruments are 12.2~ks, 23.7~ks,
11.2~ks, and 10.7~ks, respectively.
LECS and MECS data were extracted centered on the position of 4U\,1700-37 
using radii of 8\arcmin\ and 4\arcmin, respectively. 
Background subtraction in the imaging instruments
was performed using standard files, but is not critical for such a bright
source. 
Background subtraction in the non-imaging instruments was 
carried out using data from the offset intervals.
The background subtracted count rates in the LECS, MECS, HPGSPC
and PDS were 5.3, 18.5, 40.5, and 27.0~s$^{-1}$, respectively. 

\subsection{Spectral fits}

The 4U\,1700-37 spectrum was investigated by simultaneously
fitting data from all the NFI.
The LECS and MECS spectra were rebinned to oversample the full
width half maximum of the energy resolution by
a factor 3 and to have additionally a minimum of 20 counts 
per bin to allow use of the $\chi^2$ statistic. Data
was selected in the energy ranges
0.5--5.0~keV (LECS) and 1.8--10.0~keV (MECS) 
where the instrument responses are well determined. The HPGSPC and PDS
data were rebinned using standard procedures in the energy ranges 7--40~keV 
and
15--200~keV, respectively.
The photoelectric absorption
cross sections of Morrison \& McCammon (1983) and the
solar abundances of Anders \& Grevesse (1989) are used throughout.

The spectrum was first fit with an absorbed power-law model, 
including an iron line, but with no
cutoff or soft component. 
The iron line energy, ${\rm E_{Fe}}$,
was fixed at 6.5 keV, but the width, ${\rm \sigma_{Fe}}$, 
and normalization (and all other
parameters) were allowed to vary. Unsurprisingly, given the simplicity
of the model, only a very poor fit with a reduced $\chi^{2}$ 
of 47.0 for 280 degrees of
freedom (dof) was obtained. Matters are significantly 
improved if a high energy
cutoff and a low energy thermal bremsstrahlung component
are included, resulting in a reduced $\chi^{2}$ of 2.00 for 276 dof. 
The best-fit thermal bremsstrahlung
component (see Haberl et al. 1994) has a kT 
of $0.2 \pm 0.1$~keV.
Both continuum components suffer low-energy absorption, ${\rm N_H}$, of 
$(5.1 \pm 0.2) \times 10^{20}$~atom~cm$^{-2}$.
The line width was consistent with a narrow line and so was fixed
at a value (0.1~keV) much smaller than the instrumental resolution.
This gives an equivalent width, EW, of $120 \pm 20$~eV for the iron line.
The functional form 
of the cutoff above energy ${\rm E_{cut}}$ 
is ${\rm \exp[(E_{cut} - E)/E_{fold}]} $, where 
${\rm E_{cut}}$ and ${\rm E_{fold}}$ are the cutoff and folding energies,
respectively.
The count rate spectrum for this model is shown in
the left panels of Fig.~1, together with the contributions to
$\chi$$^{2}$.

A significant contribution to the existing poor fit quality is residual
structure in the PDS spectrum at $\sim$37~keV (see the left hand panels
of Fig.~1). Note that the same structure is not evident in the HPGSPC
spectrum. Since this feature is
reminiscent of a cyclotron resonance line, we added a 
single cyclotron component 
(the {\sc cyclabs} model in {\sc xspec})
to the model and re-fitted the data. 
The fit quality improves sharply, with
a reduced $\chi$$^{2}$ of 1.42 for 273 dof.
This improved fit is shown in the right hand panels of Fig.~1.
The best-fit values of ${\rm E_{cut}}$ and ${\rm E_{fold}}$ are
$5.9 \pm 0.2$~keV and $23.9 \pm 0.5$~keV, respectively. These values are 
consistent with those of $6.6 \pm 0.7$~keV
and $21.1 ^{+4.3} _{-3.3}$~keV found by Haberl et al. (1989) 
using EXOSAT data. 
Haberl \& Day (1992) determined a slightly higher cutoff
energy using {\it Ginga}, measuring
${\rm E_{cut}}$ = 7.6~keV, and ${\rm E_{fold}}$ = 19.5~keV, with no
uncertainties quoted. Nonetheless, given that these observations were made
with different missions at different epochs, the broad agreement 
is gratifying and shows that the cutoff is a necessary component in
the spectrum.

\begin{figure*}
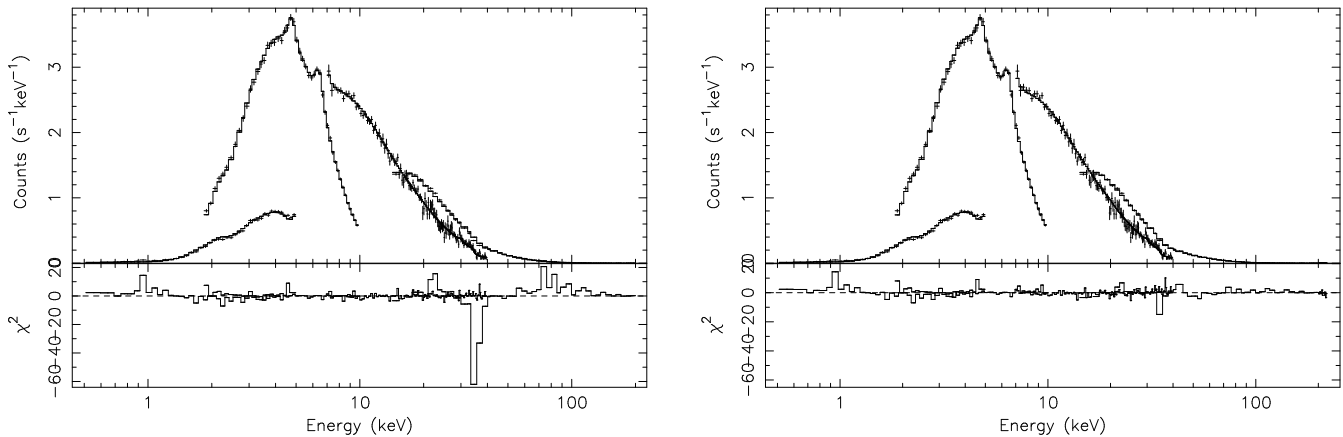

\begin{center}
\hbox{
\epsfig{file=1700_spec13.ps, angle=-90, width=8.5cm, clip=}
\hspace{0.5cm}
\epsfig{file=1700_spec23.ps, angle=-90, width=8.5cm, clip=}}
\caption{The NFI 0.5--200 keV spectrum of 4U\,1700-37.
The left panels show the fit to an absorbed 
power-law and low-energy thermal bremsstrahlung model
together with an iron emission line at 6.5~keV and an exponential cutoff 
beginning near 6~keV. In the right panels a broad cyclotron absorption
feature at $\sim$37~keV is included in the model.
The contributions to $\chi^{2}$ are plotted in
the lower panels}
\label{fig:spectrum}
\end{center}
\end{figure*}

\begin{table}
\caption[]{Best-Fit BeppoSAX spectral parameters. Norm${\rm _{PL}}$ and
Norm${\rm _{brem}}$ are the power-law 
and thermal bremsstrahlung
normalizations in units of Photon keV$^{-1}$~cm$^{-2}$~s$^{-1}$.
A${\rm _{FE}}$ is the total number of iron line photons in units of
cm$^{-2}$~s$^{-1}$. All uncertainties are quoted at 90\% confidence.
The values without uncertainties were held fixed in the fits}
\begin{flushleft}
\begin{tabular}{ll}
\hline\noalign{\smallskip}
Parameter & Value \\
\noalign{\smallskip\hrule\smallskip}
${\rm N_{H}}$ (10$^{22}$~atom~cm$^{-2}$) & 5.1 $\pm$ 0.2 \\
$\alpha$                                 & $1.07 ^{+0.02} _{-0.03}$ \\
Norm${\rm _{PL}}$                        & $0.17 \pm 0.08$ \\
${\rm E_{FE}}$ (keV)                     & 6.5 \\
${\rm \sigma_{FE}}$   (keV)              & 0.1 \\
A${\rm _{FE}}$                           &$(2.7 \pm 0.2) \times 10^{-3}$ \\
EW${\rm _{FE}}$  (keV)                   & $0.120 \pm 0.020$ \\
kT (keV)                                 & 0.2 $\pm$ 0.1 \\
Norm${\rm _{brem}}$                      & 180$^{+480} _{-120}$\\
${\rm E_{cut}}$ (keV)                    & 5.9 $\pm$ 0.2 \\
${\rm E_{fold}}$ (keV)                   & 23.9 $\pm$ 0.5 \\
${\rm E_{cyc}}$ (keV)                    & 36.6 $\pm$ 1.0 \\
${\rm \sigma_{cyc}}$ (keV)                 & 11$^{+5}_{-3}$ \\
\noalign{\smallskip}
\hline
\end{tabular}
\end{flushleft}
\label{tab:bestfit}
\end{table}

The best-fit
energy, ${\rm E_{cyc}}$, and width, ${\rm \sigma_{cyc}}$, 
of the feature are $36.6 \pm 1.0$~keV and $11 ^{+5}_{-3}$~keV, respectively.
This width is a factor $\sim$2.5 broader then expected from
the correlation between
these parameters observed by
Dal Fiume et al. (1998) for five other massive X-ray binaries.
This suggests that the feature may arise through incorrect modeling
of the continuum.
This is supported by the lack of detection in the HPGSPC.
Variations in PDS performance, which could produce 
a similar spectral feature if the background and source spectra had slightly
different gains, are however excluded.
The feature is still present if the continuum is modeled with a
broken power-law together with a high-energy 
cutoff, as suggested by Dal Fiume et al. (1998). 
Following Mihara (1995) and Dal Fiume et al. (1998), 
we also generated a Crab ratio
spectrum, by dividing the PDS data by those obtained from a 
PDS observation of the Crab Nebula, which
is time-invariant and has a smoothly varying E$^{-2.1}$ continuum
in this energy range. This technique
is useful in minimizing instrumental and model dependent effects.
While
the Crab ratio spectrum indicates some structure at $\sim$37~keV, it
does not show the pronounced deficit which is the signature of the
cyclotron features discussed in Dal Fiume et al. (1998). 
We conclude 
that while the suggestion of a cyclotron line is intriguing, and results
in a significantly lower value of $\chi^{2}$, we cannot
exclude the possibility that the feature is an artefact of the fitting
process.

Further improvements in fit quality require a better description
of the spectral region covered by the LECS, since the main
remaining contribution to $\chi^{2}$ is a feature at $\sim$1~keV 
(see Fig.~1).
This feature has not been seen in broad-band spectra of
4U\,1700-37 before, since the {\it Ginga} observations did not cover the
energy region below 1.5~keV, but it may also be evident in the residuals
of the ROSAT PSPC spectrum presented in Haberl et al. (1994).
Since Rubin et al. (1996) show that the BATSE 
spectrum above 20~keV is
consistent with a thermal bremsstrahlung model, we
examined the PDS data separately to check for consistency with this earlier
study.
The result of Rubin et al. (1996) is confirmed, with the PDS spectrum
being described by a thermal 
bremsstrahlung with kT = $25.9 \pm 0.3$~keV, in
agreement with the temperature of 25~keV derived from the
BATSE data, for a reduced $\chi^{2}$ of 2.16 for 46~dof.
The addition of a broad cyclotron line at $\sim$40~keV 
improves the fit significantly giving a reduced
$\chi^{2}$ of 1.03 for 43~dof. 

\subsection{The nature of the compact object in 4U\,1700-37}

The broadband BeppoSAX spectrum presented here is very
similar to the X-ray spectrum of an accreting pulsar. The underlying
shape is an absorbed power-law sharply 
modified $\approxgt$6.0~keV by a cutoff. Although the photon index
of $\sim$1.0 is rather higher than the average reported by
previous authors, it lies within the range of observed variation
exhibited by 
4U\,1700-37, and is at the upper range of the values for various
pulsars summarized in White et al. (1983). 

In what respects does the 4U\,1700-37 spectrum differ from, or resemble 
that, of a black hole? Four
diagnostics of black hole candidates (none
of them individually conclusive) are listed in Tanaka \& Lewin (1995).
These are: (1) ultrasoft spectra when luminous, (2) high-energy power-law
tails, (3) bimodal spectral states, and (4) millisecond variability and
flickering in the hard state. 4U\,1700-37 possesses none of these
attributes. In particular,
steep, cutoff spectra (${\rm E_{cut}} \approxlt$20~keV)  
are a feature of neutron stars rather than black holes (White et al. 1988),
while black holes (at least in outburst) generally 
possess hard tails, emitting
significant flux $\approxgt$100~keV (see e.g., Ballet et al. 1994). 4U\,1700-37
resembles neither the ultrasoft ``high-state'', nor the hard power-law 
``low-state''
spectrum of the canonical black hole candidate Cyg X-1.

\section{Discussion}

We have demonstrated that the broadband 
X-ray spectrum of 4U\,1700-37 is
qualitatively similar to that of other accreting neutron star
X-ray pulsars.  A feature at $\sim$37~keV may be modeled as a broad cyclotron
absorption line, but its presence is uncertain due to uncertainties in
continuum modeling and instrument performance. 
Such features have only been reported in 
about
half of the known X-ray pulsars -- and in some cases only at marginal
significance.  The lack of pulsations remains
puzzling, and may suggest (as has been argued by previous authors) 
that the magnetic field is intrinsically
weak or aligned with the rotation axis. The neutron star's
progenitor may have been very massive, suggesting that there may be
a range of masses over which progenitors form both black holes
and neutron stars, or that black holes are formed only through
a limited range of progenitor masses, above which the remnants
are again  neutron stars.

\begin{acknowledgements}
The BeppoSAX satellite is a joint Italian-Dutch programme.
We thank Fabio Favata, Matteo Guainazzi, and Tim Oosterbroek for
helpful discussions. We also thank the referee, Frank Haberl,
for suggestions which helped improve the paper.
\end{acknowledgements}

\vfill
\eject

\end{document}